\documentclass[aps,prl,twocolumn,amsmath,amssymb,amsbsy,superscriptaddress,floatfix,showpacs]{revtex4}
\usepackage[dvips]{graphicx}
\begin{document}
\title{Thermodynamics of a Fermi liquid in a magnetic field}
\author{Joseph Betouras}
\affiliation{ Max--Planck--Institut f{\"u}r Physik komplexer
Systeme, N{\"o}thnitzer Str. 38, 01187 Dresden, Germany. }
\author{Dmitry Efremov}
\affiliation{ Lehstruhl f{\"u}r Festk{\"o}pertheorie, TU Dresden
Institut f{\"u}r Theoretische Physik 01062 Dresden, Germany. }
\author{Andrey Chubukov}
\affiliation{ Department of Physics and Condensed Matter Theory Center,
 University of Maryland, College Park MD 20742-4111, USA. }
\date{\today}
\begin{abstract}
  We extend previous calculations of the non-analytic terms 
 in the spin susceptibility $\chi_s (T)$ 
 and the specific heat $C(T)$ to systems in a  magnetic field.
Without a field,  $\chi_s (T)$ and $C(T)/T$ are linear in $T$ in $2D$, while in
 $3D$, $\chi_s (T) \propto T^2$ and $C(T)/T \propto T^2 \log T$. 
 We show that in a magnetic field, the linear in $T$ terms in 2D become
  scaling functions of $\mu_B H/T$. We present explicit
 expressions for these functions and show that at high fields, 
$\mu_B H \gg T$, $\chi_s (T,H)$ scales as $|H|$.
We also show that in 3D, $\chi_s (T, H)$ becomes non-analytic in a
 field and at high fields scales as $H^2 \log |H|$.
 \end{abstract}
\maketitle

Landau Fermi liquid theory  \cite{landau} provides the
basis for our present understanding of correlated electronic
systems. The  theory predicts that, in any Fermi liquid, 
the spin susceptibility
$\chi_s (T)$ and the specific heat coefficient $\gamma (T) =C(T)/T$ 
tend to a constant at $T \to 0$ \cite{landau,AGD}. Later,  Landau theory has
been extended to include the leading temperature dependence
of $\chi_s (T)$ and $\gamma (T)$ which turn out to be non-analytic in 
dimensions $D\leq 3$ \cite{doniach,{amit},{pethick},{belitz},{millis},
{bedell},baranov,{andrey1},dassarma,{andrey2},{aleiner},{galitski}}.
Like the zero-temperature terms, the thermal corrections 
 come from fermions in the 
immediate vicinity of the Fermi surface. In 2D systems, both 
$\chi_s (T)$ and $\gamma (T)$ are linear in $T$~\cite{bedell,baranov,
millis,andrey1}
 and the coefficients are
expressed  in terms of charge and spin components of the scattering amplitude
at the scattering angle $\theta = \pi$ \cite{andrey2,aleiner}. 
In 3D, $\chi_s (T)$ is quadratic in 
$T$, i.e., is analytic \cite{{amit},{pethick},{belitz}}, 
while $\gamma (T)$ is non-analytic and 
scales as $T^2 \log T$ \cite{doniach,amit,pethick}.    

In this communication, we extend  previous works to 
systems in a
magnetic field $H$. We consider $S=1/2$ 
charge-less fermions (like $^3He$ atoms) for which  
the magnetic field adds spin-dependent 
Zeeman term $\pm \mu_B H$ to the fermionic  
dispersion.  We  show that, in the presence of a field, 
$\Delta \chi_s(T,H) = \chi_s (T,H) - \chi_s (0,0)$ and $\Delta \gamma (T,H) = 
\gamma (T,H) - \gamma (0,0)$  become scaling functions of $\mu_B H/T$:
$\Delta \chi_s (T,H) = \Delta \chi_s (T,0) f_{\chi} (\mu_B H/T)$, 
~~$\Delta \gamma (T,H) = \Delta \gamma (T,0) f_{\gamma} (\mu_B H/T)$.
We present the  expressions for these functions to second order in the
interaction potential $U$. For 2D systems, we show that at $\mu_B H \gg T$ (but still, $\mu_B
H \ll E_F$), $\Delta \chi_s (T,H)$ scales as $|H|$ and 
weakly depends on $T$.
In the same field range, $\delta \gamma (T,H)$ is still linear in $T$, but 
the prefactor is different from that at $H=0$. 
For 3D systems,
  we show that $\Delta \chi_s (T,H)$ becomes non-analytic at a non-zero 
$H$. The non-analytic term in $\Delta \chi_s (T,H)$ scales as 
$H^2 \log [max (\mu_B H, T)/E_F]$. The specific heat coefficient 
$\gamma (H,T)$ in a field still scales as  $T^2 \log T$, but, like in 2D,
 the prefactor changes between $H=0$ and $\mu_B H \gg T$. 

The analysis of the behavior of $\Delta \chi_s(T,H)$ and $\Delta \gamma (T,H)$
in a magnetic field may be useful for  experimental verifications of the
non-analytic behavior of thermodynamic variables.  It is
more straightforward to analyze the dependence on the magnetic field
rather than the dependence on the temperature. In particular, recent  
measurements of the  temperature dependence  of the spin susceptibility
in Si inversion layers~\cite{R1} didn't yield  conclusive results
on whether the $T$ dependence of $\chi_s (T)$ is indeed linear,
as some temperature dependence inevitably comes from  spins
on the substrate. We propose to measure the field dependence of the
spin susceptibility at a given $T$ and use our scaling functions 
to fit the data. 
 
The point of departure for our calculations 
is the Luttiger-Ward expression for the thermodynamic 
potential. To simplify the presentation, we assume that the interaction 
potential $U(q)$ is independent on $q$. We restore the momentum dependence
of $U(q)$ in the final formulas.  
To  second-order in $U$, the thermodynamic potential
is given by
\begin{equation}
\Phi = \Phi_0 -\frac{U^2}{2} T \sum_n \int_q \frac{d^dq}{(2\pi)^d}
\Pi^{\uparrow \uparrow}(\vec{q},\Omega_n, T)
\Pi^{\downarrow \downarrow}(\vec{q},
\Omega_n, T),
\label{1}
\end{equation}

\noindent where $\Phi_0$ is the thermodynamic potential for free fermions, 
 and  $\Pi^{\uparrow \uparrow} (\vec{q},\Omega_n, T)$ and 
$\Pi^{\downarrow \downarrow}(\vec{q},
\Omega_n, T)$ are the particle-hole
 bubbles composed of fermions with spin up or spin down, respectively.

Previous studies of the spin susceptibility and the specific heat in a 
zero magnetic field established that the  
non-analytic temperature behavior of $\Delta \chi_s (T)$  and 
$\Delta \gamma (T)$  originates from the 
non-analyticity of the polarization operator either near 
$q=0$ (Landau damping) \cite{doniach,amit,pethick,{andrey1}}  or near
$q=2k_F$ (a dynamic Kohn anomaly) \cite{belitz,baranov,{millis},
{andrey1},{aleiner},galitski}. 
The $2k_F$ non-analyticity contributes to  the spin susceptibility
and the specific heat, while the $q=0$ non-analyticity 
only contributes to the non-analyticity in the specific heat.
This can be easily understood as the non-analytic term in the 
zero field 
spin susceptibility $\Delta \chi_s (T)$  describes a singular response
 to an infinitesimally small magnetic field. A magnetic field 
splits Fermi momentum $k_F$ into 
$k^{\uparrow}_F$ and $k^{\downarrow}_F$. The small $q$ form of 
  $\Pi^{\uparrow\uparrow}(\vec{q},\Omega_n, T)$ and
  $\Pi^{\downarrow\downarrow}(\vec{q},\Omega_n, T)$ is
unaffected by this splitting, up to terms of order $(\mu_B H/E_F)^2$,
hence the response to the infinitesimal field
must be analytic in $T$.
At the same time,
 singular $2k_F$ contribution to $\Phi (T)$ at zero field originates from 
the fact that the two polarization operators in Eq. (\ref{1}) are 
non-analytic at the same $q=2k_F$. In a field the singularities in spin-up and
 spin-down polarization operators  occur at different $q=2k_F^{\uparrow}$ and $q= 2k_F^{\downarrow}$. Accordingly, a  magnetic field regularizes $2k_F$ non-analyticity in the thermodynamic potential, 
but for a price that the linear response to the field, i.e. the 
 spin susceptibility $\Delta \chi (T,H=0)$, becomes non-analytic.    

Our  goal is to analyze the forms of the susceptibility and the
 specific heat at a finite $H$, i.e., beyond the linear response theory.
We consider the fields for which $\mu_B H$ is comparable to $T$, but still
$\mu_B H \ll E_F$. For these fields, the non-analytic
contribution to $\Phi$ from small $q$ are unaffected by the field . However the $2k_F$ contribution is  field
dependent and evolves at $\mu_B H \sim T$. 
\begin{figure}[tbp]
\centering
\includegraphics[width=1.0\columnwidth]{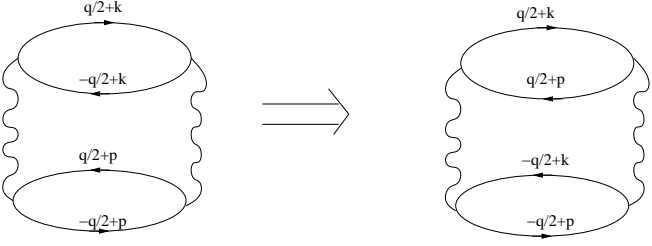}
\caption{The  diagram for $\Delta\Phi_{2k_F}$, and the
 trick to compute it. As the non-analytic part of $\Delta \Phi_{2k_F}$ 
 comes from small $k$ and $p$, it can be re-expressed as a
 product of two bubbles $\Pi^{\uparrow \downarrow} (q', \omega)$ 
with small momentum transfer $\vec{q'} = \vec{k} - \vec{p}$.}
\label{fig1}
\end{figure}  

The calculation of $\Delta \Phi = \Phi - \Phi_0$ is somewhat tricky.
In principle,  all one has to do is to evaluate 
  particle-hole bubbles for fermions
 with up and down spins at a finite $T$, substitute the results  
into Eq. (\ref{1}), 
integrate over momentum $q$ and sum over Matsubara frequencies $\Omega_n$.
In practice, however,
this computation is easy to perform only for small $q$ part as 
for  $q \ll k_F$, the non-analytic part of the 
polarization bubble is associated
with the Landau damping, which does not depend on $T$, apart from regular
$(T/E_F)^2$ corrections. Accordingly, one can safely use the known
 analytical  forms
 of $\Pi (q, \Omega)$ at $T=0$.  
For $q$ near $2k_F$, non-analytic terms in 
$\Pi^{\uparrow \uparrow} (\vec{q},\Omega_n, T)$ and 
$\Pi^{\downarrow \downarrow}(\vec{q},
\Omega_n, T)$ contain scaling functions of $T/\omega$,  which 
 are only available  in  integral forms \cite{millis}. This substantially complicates  direct calculation of the $2k_F$ term.
 There exists,  however,  a way to compute the $2k_F$ term, which 
 avoids dealing with the $2k_F$ polarization bubbles at a finite $T$. 
This method explores
the fact that only the non-analytic parts
$\Pi^{\uparrow \uparrow} (\vec{q},\Omega_n, T)$ and 
$\Pi^{\downarrow \downarrow}(\vec{q}, \Omega_n, T)$ for $q$ near $2k_F$ 
contribute the non-analyticity in the thermodynamic potential.
Earlier works have demonstrated that the $2k_F$ non-analyticity
in $\Pi(\vec{q},\Omega_n, T)$  comes 
from fermions in the particle-hole bubble with 
momenta near $\pm {\vec q}/2$ \cite{belitz,andrey2}. 
This implies that, out of four fermions in the 
 second order, two-bubble diagram for the thermodynamic potential 
in Fig. 1, two
fermions with  opposite spins have momenta near ${\vec q}/2$, while the 
other two
fermions have momenta near $-{\vec q}/2$. 
Then the $2k_F$ part of the $\Delta \Phi$ can be re-written as  
the integral over small ${\vec k}$ and small
${\vec p}$ of
\begin{eqnarray}
&& \Delta \Phi_{2k_F} =-\frac{U^2}{2} \sum_{\omega_m,\omega^\prime_m,\omega_m^{\prime \prime} } \int d^2 q \int d^2 k d^2 p \nonumber \\
&& G^{\uparrow}(\vec{q}/2+\vec{k}, \omega_m+\omega_m') G^{\downarrow}(\vec{q}/2 , \omega^\prime_m) \times \nonumber \\
&& G^{\uparrow}(-\vec{q}/2 +{\vec k} , \omega_m + \omega_m^{\prime \prime})
  G^{\downarrow}(-\vec{q}/2+\vec{p},  \omega_m^{\prime \prime})
\label{2}
\end{eqnarray}
or, equivalently, as  
\begin{equation}
\Delta \Phi_{2k_F}=  -\frac{U^2}{2} \sum_n \int \frac{d^d q'}{(2\pi)^d}
\left[\Pi^{\uparrow \downarrow}(\vec{q'},\Omega_n, T)\right]^2 ,
\label{3}
\end{equation}
\noindent where the integration is confined to small $\vec{q'} = \vec{k}-\vec{p}$. In other words, the
non-analytic $2k_F$ contribution to
the thermodynamic potential can be re-expressed in 
terms of the particle-hole bubble for fermions with opposite spins and 
a {\it small} momentum transfer. The
non-analytic term in $\Pi$ at small $\vec{q'}$
does not depend on temperature (apart from irrelevant corrections), hence
$\Pi^{\uparrow \downarrow}(\vec{q'},\Omega_n, T)$ can be safely approximated by
$\Pi^{\uparrow \downarrow}(\vec{q'},\Omega_n, 0)$. 
 At the same time, 
the polarization bubble $\Pi^{\uparrow \downarrow}(\vec{q},\Omega_n, 0)$
strongly depends on the magnetic field (contrary to 
$\Pi^{\uparrow \uparrow}(\vec{q},\Omega_n, 0)$), and this gives rise to the
 scaling dependence on $\mu_B H/T$. 

Combining the $q=0$ and $2k_F$ contributions, we  obtain
for the thermodynamic potential 
\begin{eqnarray}
&&\Delta \Phi = -\frac{U^2}{2} \sum_n \int \frac{d^d q}{(2\pi)^d} \nonumber \\
&&\left[\left( \Pi^{\uparrow \downarrow}(\vec{q},\Omega_n, 0) \right)^2
+ \Pi^{\uparrow \uparrow}(\vec{q},\Omega_n, 0)
\Pi^{\downarrow \downarrow}(\vec{q},\Omega_n, 0)\right] ,
\label{3_1}
\end{eqnarray}
where the integration involves only small $q$.

We next proceed separately with $2D$ and $3D$ cases. 
In 2D we have \begin{eqnarray}
&&\Pi^{\uparrow \downarrow}(\vec{q},\Omega_n, T) =
\frac{m}{2 \pi} \frac{|\Omega_n|}{\sqrt{(\Omega_n - i \; \delta
\mu)^2 + (\upsilon_{F}q)^2}} + ... \nonumber \\
&&\Pi^{\uparrow \uparrow}(\vec{q},\Omega_n, T) =
 \frac{m}{2 \pi} \frac{|\Omega_n|}{\sqrt{\Omega^2_n + (\upsilon_{F}q)^2}} + ...    ,
\label{4}
\end{eqnarray}
where dots stand for analytic terms, expandable in powers of $\Omega^2_n$ or $q^2$, and $\delta \mu =
\mu_{\uparrow}-\mu_{\downarrow}=2 \mu_B H$. 
Substituting Eq. (\ref{4}) into Eq. (\ref{3_1}) and integrating 
 explicitly over momentum $q$ we obtain 
\begin{equation}
\Delta \Phi =  \left(\frac{m}{2 \pi}\right)^2
\frac{ U^2 T}{8 \pi v_{F}^2} \sum_n {\Omega_n}^2 \log \left[ \frac{(\Omega_n - 2 i \mu_B H)^2  \Omega^2_n}{E^4_F} \right]. \label{5}
\end{equation}
Differentiating Eq. (\ref{5}) with respect to $H$, we obtain
$$
\Delta M = - \frac{\partial \Delta \Phi}{\partial H} =
\frac{{\mu_B}^4 m^4 U^2 H^3}{{\pi}^3 k_{F}^2} T \sum_n
\frac{1}{{\Omega_n}^2 + (2 \mu_B H)^2}
$$
The sum over  Matsubara frequencies can be easily evaluated and yields 
\begin{eqnarray}
\nonumber \Delta M &=& \frac{ \mu_B m^4 U^2 }{4 \pi^3 k_{F}^2}
T^2 \left[ \left(\frac{\mu_B
H}{T} \right)^2 \coth(\frac{\mu_B H}{ T}) \right] \\
&=& \mu_B A T^2 x^2 \left[ 1 + 2 n_B(2x)\right]  ,
\label{7}
\end{eqnarray}
 where 
\begin{equation}
A= \frac{  m^4 U^2}{4 \pi^3 k_{F}^2}, ~~~x=\frac{\mu_B H}{ T}.
\label{8}
\end{equation}
We see from Eq. (\ref{7}) that $\Delta M$ increases in a field by
two reasons. First, the field leads to a 
finite magnetization at $T=0$, and second, a finite field
 populates the system with  spin waves precessing at the 
energy $\mu_B H$.
 Differentiating (\ref{7}) again over $H$, we obtain the spin susceptibility 
in the form 
\begin{equation}
\Delta \chi(T,H)= \chi (T,H) - \chi (0,0) = 
\mu_B^2 A  T f_{\chi} (x)     ,
\label{9}
\end{equation}
where 
\begin{equation}
f_\chi (x) = \frac{x}{\sinh^2(x)} \left[\sinh(2x) - x \right]    .
\label{10}
\end{equation}
For vanishing $H$, i.e., at $x \rightarrow 0$, $f_\chi (0) =1$, and 
\begin{equation}
\Delta \chi(T,H)= \chi (T,0) - \chi (0,0) =  \mu_B^2 A  T      .
\label{10'}
\end{equation}
This coincides with the earlier result \cite{andrey1}.
In the opposite limit of large $x$, $f_{\chi}(x) \approx 2x$, and 
\begin{equation}
\Delta \chi(T,H)=  2 \mu_B^2 A  T |x|  = 2 \mu_B^3 A |H|     .
\label{10''}
\end{equation}
We see that at high fields, the spin susceptibility scales as $|H|$,
 i.e., is non-analytic in $H$.
\begin{figure}[tbp]
\centering
\includegraphics[width=\columnwidth]{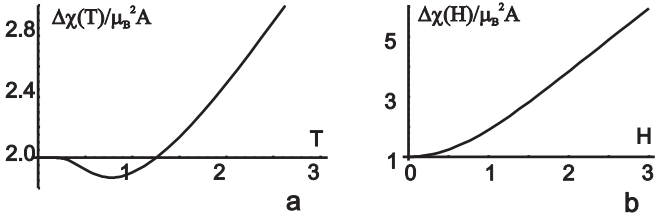} 
\caption{$\Delta \chi$ as a function
 of temperature at fixed magnetic field (a) and 
  as a function of magnetic field
at fixed temperature (b). In the left panel,
 $T $ and $\Delta \chi/ \mu_B^2 A $ are in units of $\mu_B H$ 
($A$ is defined in the text). In the right panel 
 $\mu_B H $ and $\Delta \chi/ \mu_B^2 A $ are in 
units of $T$. In these units, 
$\Delta \chi /\mu^2_B A = f_{\chi}(\mu_B H )$).            
\label{fig2}}
\end{figure}

In Fig. 2 we plot the susceptibility as a function of temperature at 
a given $H$, and as a function of the magnetic field at a given $T$.
Note that at a finite $H$, the Bose term in Eq. (\ref{8}) 
gives rise to a negative derivative of $ \partial \Delta \chi/ \partial T$. 
This in turn gives rise to a shallow minimum in the 
temperature dependence of $\Delta \chi (T, H)$.
 
The specific heat 
$\Delta \gamma (T,H) = \gamma (T,H) - \gamma (0,0)$
is obtained by differentiating Eq. (\ref{5}) twice over $T$.
At $H=0$, $\Delta \gamma(T,H)= -6AT \zeta(3)$~\cite{andrey1,andrey2,aleiner}.
At a finite $H$,  
\begin{eqnarray}
&&\Delta \gamma(T,H)= -6AT \zeta(3) + 
2A T \int_0^{\mu_B H/T}  \frac{dx x^3} {\sinh^3x} \times \nonumber \\
&&~\left( x \cosh
x -\sinh x\right) = - A T f_\gamma (x)    ,
\label{11}
\end{eqnarray}
where 
\begin{eqnarray}
&&f_\gamma (x) =3 \left(
Li_{3}(e^{- 2 x}) + 2 x Li_{2}(e^{-2x}) -
2 x^2 \log(1-e^{-2x}) \right) \nonumber \\
&&+  6 \zeta (3) - 2 x^3  +4 x^3 \coth x -
x^3 \frac{1}{\sinh^2 x }\left( \sinh 2x - x \right) \nonumber    
\end{eqnarray}
and  $Li$ are polylogarithmic functions. At  $x \ll 1$, $f_\gamma (x) \approx 6 \zeta (3)- \frac{x^4}{6}$ and
$$
\Delta \gamma (\frac{\mu_B H}{T} \ll 1) \approx  - AT \left(6 \zeta (3) -
\frac{1}{6} \left(\frac{\mu_B \ H}{T}\right)^4 \right).
$$
 In the opposite limit of $x \gg 1$, 
$f_\gamma (x) = 3 \zeta (3) + 4 x^4 e^{-2x}$, and
$$
\Delta \gamma (\frac{\mu_B H}{T} \gg 1)  \approx
 - A T \left(3 \zeta (3) + 4\left( \frac{\mu_B H}{T}\right)^4 
e^{-2 \mu_B H/T }\right).
$$
 We see that in both limits
the temperature dependence of the specific heat is linear in $T$, but
the prefactor changes by a factor of $2$ between small and high fields. 
This result could be anticipated as a high magnetic field eliminates the
non-analyticity in  the polarization bubble 
$\Pi^{\uparrow \downarrow}(\vec{q},\Omega_n,)$, such that only the 
second term in Eq. (\ref{3_1}) contributes to 
the $T$ term in $\Delta \gamma (T,H)$.

\begin{figure}[tbp]
\centering
\includegraphics[width=\columnwidth]{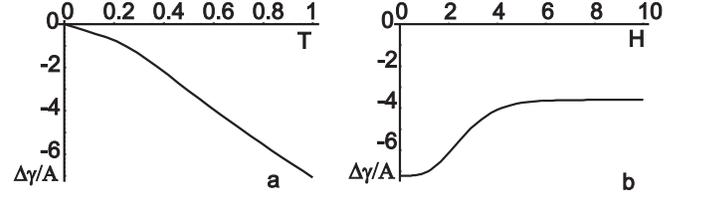}
\caption{$\Delta \gamma$ as a function  of temperature
at fixed magnetic field (a) and 
 as a function of  magnetic field
at fixed temperature (b). 
In the left panel,
 $T $ and $\Delta \chi/ \mu_B^2 A $ are in units of $\mu_B H$. 
In the right panel  $\mu_B H $ and $\Delta \chi/ \mu_B^2 A $ are in 
units of $T$. In these units, 
 $\Delta \gamma / A = f_{\gamma}(\mu_B H ) $.
} \label{fig3}
\end{figure}

The extension of the above results 
to an arbitrary $U(q)$ is straightforward. 
For the susceptibility, the
 prefactor in Eq. (\ref{9}), contains $U(2k_F)$ instead of
 $U$\cite{belitz,millis,andrey1}. 
For the specific heat coefficient, we have, instead of (\ref{11})
 \begin{eqnarray}
&&\Delta \gamma(T,H)=  -\frac{3 \zeta (3) m^4}{2 \pi^3 k_{F}^2} T \times
\left[ \left(U(0) -\frac{1}{2} U(2k_F)\right)^2 \right.
\nonumber \\
&& \left.
+ \frac{U^2(2k_F)}{4} \left(1 + 
2 \frac{ f_\gamma (x)-3 \zeta (3)}{3\zeta(3)} \right)\right] . 
\label{16}
\end{eqnarray} 
The combinations $U(0) - 1/2 U(2k_F)$ and $-1/2 U(2k_F)$ are
charge and spin components of the scattering amplitude $A(\pi)$, respectively.
At large 
$x$, $f_\gamma (x) \approx 3\zeta (3)$, and the last term in the r.h.s. of
(\ref{16})  vanishes. 
This obviously implies that at a large
field, only the charge the longitudinal spin components of the 
scattering amplitude 
 contribute to $\Delta \gamma (T)$.

We next consider the $3D$ case. 
The polarization operators at small $q$ are  given by \cite{AGD}:
\begin{eqnarray}
&&\Pi^{\uparrow \downarrow}(\vec{q},\Omega_n, 0) =
 \frac{m k_F}{2 \pi^2} \frac{\Omega_n}{v_F q} \arctan \frac{
(\Omega_n - i \; \delta \mu)}{v_F q} + ... \nonumber \\
&&\Pi^{\uparrow \uparrow}(\vec{q},\Omega_n, T) =
 \frac{m k_F}{2 \pi^2} \frac{\Omega_n}{v_F q} \arctan \frac{\Omega_n}{v_F q}
 + ...
\label{4_1}
\end{eqnarray}
As before, 
dots stand for analytic terms, expandable in powers of 
$\Omega^2_n$ or $q^2$, and $\delta \mu =
\mu_{\uparrow}-\mu_{\downarrow}=2 \mu_B H$.  
Differentiating the thermodynamic potential, Eq. (\ref{3_1}), 
with respect to $H$, we obtain
$$
\Delta M = - \frac{\partial \Delta \Phi}{\partial H} =
- \frac{{\mu_B} (m U  k_F)^2}{4 \pi^5 v^3_F} T \sum_n
{\Omega_n}^2  \arctan \frac{2 \mu_B H}{|\Omega_n|}        .
$$
Differentiating further with respect to $H$, we obtain 
\begin{eqnarray}
&&\Delta \chi_s (T,H)= - \frac{{\mu_B} (m U  k_F)^2}{2 \pi^5 v^3_F} \nonumber \\&&
 \left[ T \sum_{n=1}^M \Omega_n + 4 (\mu_B H)^2 T \sum_{n=1}^M  
\frac{\Omega_n}{\Omega^2_n + 4 \mu^2_B H^2}\right]       ,
\label{17}
\end{eqnarray}
where $M \sim E_F/T$ is the upper cutoff in the  summation over frequency.
The first term in the r.h.s of Eq. (\ref{17}) is the susceptibility at
 zero field. By power counting, one might expect the
 spin susceptibility $\chi_s (T)$ in 3D to scale as $T^2 \log T$. 
 However, the Matsubara sum $T \sum_{n=1}^M \Omega_n$ only contains  a T-independent term, of order $E^2_F$, and a  
 term $-(1/6) \pi T^2$.
This last  term is universal, but it is
 analytic in $T$. As a result, $\Delta \chi_s (T,0) \propto T^2$ is 
 analytic and  essentially irrelevant as
 the analytic in $T$ contributions to $\chi_s (T)$ are already present in the
 Lindhard function for free fermions. The absence of the non-analytic 
temperature correction to the spin susceptibility in $3D$ 
was first noticed in Ref. \cite{belitz}, (see also \cite{pethick}). 
The second term in the r.h.s. of (\ref{17}) is the extra contribution
in a finite field. Evaluating the Matsubara sum we find that this 
contribution  scales as $H^2 \log \{max(T, \mu_B H)\}$. 
We see therefore 
that in a finite magnetic field,  $\chi_s (T)$ does 
indeed become non-analytic. Casting  $\Delta \chi_s (T,H)$ into the scaling 
form, we obtain 
\begin{equation}
\Delta \chi_s (T,H) = \chi_0 \left(\frac{m U  k_F}{2\pi^2}\right)^2 \left( \frac{T}{E_F}\right)^2 g \left(\frac{\mu_B H}{T}\right) ,
\label{18}
\end{equation}
where $\chi_0 = \mu^2_B k^3_F/(2\pi^2 E_F)$ is Pauli susceptibility,
 and to a  logarithmic accuracy, 
\begin{equation}
g (x) = x^2 \log \left[\frac{E_F}{T* max \{x,1\}}\right] .\end{equation}
The $H^2 \log H$ dependence of $\chi(H)$ was earlier reported by 
Misawa \cite{misawa}. However, his prefactor  is different from the 
 one we obtained.  
 
Differentiating the thermodynamic potential twice over $T$, we also
obtain field dependence of the specific heat coefficient.
 The  field dependence in 3D parallels the one for 2D systems.
Namely, at zero field, 
\begin{eqnarray}
&&\Delta \gamma (T,0) =-\frac{3}{20} \frac{(m k_F)^2}{\pi^2} \\
&&
\times \left[(U (0) -\frac{1}{2}  U (2k_{F}))^2 + \frac{3}{4} U^2 (2k_{F})\right]
 \left( \frac{T}{E_{F}}%
\right) ^{2}\ln \frac{E_{F}}{T}. 
 \label{19}  \nonumber 
\end{eqnarray}
In a finite field, the charge part is not affected, while in the spin part, 
the logarithmic factor $3 \log \frac{E_{F}}{T}$ is replaced by $\log \frac{E_{F}}{T} + 2 \log \frac{E_{F}}{max\{T,  \mu_B H\}}$.  
As a result, at $\mu_B H \gg T$, 
$\Delta \gamma (T,H)$ still behaves as $T^2 \log T$,  
but the prefactor gets smaller. 

To summarize, in this paper
we analyzed non-analytic terms in  the magnetization,
the spin susceptibility and the specific heat of  2D and 3D Fermi
liquids, placed into an external magnetic field $\mu_B H \ll E_F$. 
 We obtained  the non-analytic terms in the forms of scaling functions of 
 $\mu_B H/T$. We found that at $\mu_B H \gg T$, 
 the spin susceptibility scales as $|H|$ in
 2D  and as $H^2 \log |H|$ in 3D. The specific heat in a 
field preserves the same
temperature dependence as in the absence of a field, 
but the prefactor changes between small and large $\mu_B H/T$.

 We  thank D. Belitz, P. Fulde, 
D. Maslov and  T. Vojta for useful discussions. 
The research has been supported by NSF DMR 0240238 (A. Ch.), and 
the Visitors Program of the Max Planck Society (J.  B.)

\end{document}